\documentclass[aps,reprint,amsmath,amssymb,superscriptaddress]{revtex4-2}

\usepackage{amsmath}
\usepackage{amssymb}
\usepackage{graphicx}
\usepackage{CJK}
\usepackage{keyval}
\usepackage{dcolumn}
\usepackage{bm}
\usepackage{color}
\usepackage{txfonts}
\usepackage{verbatim}
\usepackage[rightcaption]{sidecap}

\usepackage{soul}
\usepackage{ulem}
\usepackage{booktabs}
\begin{document}

\title{Electronic and magnetic excitations in La$_3$Ni$_2$O$_7$}

\author{Xiaoyang Chen}
\thanks{Equal contributions}
\affiliation{State Key Laboratory of Surface Physics, Department of Physics, and Advanced Materials Laboratory, Fudan University, Shanghai 200438, China}

\author{Jaewon Choi}
\thanks{Equal contributions}
\affiliation{Diamond Light Source, Harwell Campus, Didcot OX11 0DE, UK}

\author{Zhicheng Jiang}
\affiliation{National Synchrotron Radiation Laboratory and School of Nuclear Science and Technology, University of Science and Technology of China, Hefei, 230026, China}

\author{Jiong Mei}
\affiliation{Beijing National Laboratory for Condensed Matter Physics and Institute of Physics,
	Chinese Academy of Sciences, Beijing 100190, China}
\affiliation{School of Physical Sciences, University of Chinese Academy of Sciences, Beijing 100190, China}

\author{Kun Jiang}
\affiliation{Beijing National Laboratory for Condensed Matter Physics and Institute of Physics,
	Chinese Academy of Sciences, Beijing 100190, China}
\affiliation{School of Physical Sciences, University of Chinese Academy of Sciences, Beijing 100190, China}

\author{Jie Li}
\affiliation{National Laboratory of Solid State Microstructures and Department of Physics, Nanjing University, Nanjing 210093, China}

\author{Stefano Agrestini}
\affiliation{Diamond Light Source, Harwell Campus, Didcot OX11 0DE, UK}

\author{Mirian Garcia-Fernandez}
\affiliation{Diamond Light Source, Harwell Campus, Didcot OX11 0DE, UK}

\author{Xing Huang}
\affiliation{Guangdong Provincial Key Laboratory of Magnetoelectric Physics and Devices, School of Physics, Sun Yat-Sen University, Guangzhou, Guangdong 510275, China}

\author{Hualei Sun}
\affiliation{School of Science, Sun Yat-Sen University, Shenzhen, Guangdong 518107, China}

\author{Dawei Shen}
\affiliation{National Synchrotron Radiation Laboratory and School of Nuclear Science and Technology, University of Science and Technology of China, Hefei, 230026, China}

\author{Meng Wang}
\affiliation{Guangdong Provincial Key Laboratory of Magnetoelectric Physics and Devices, School of Physics, Sun Yat-Sen University, Guangzhou, Guangdong 510275, China}

\author{Jiangping Hu}
\affiliation{Beijing National Laboratory for Condensed Matter Physics and Institute of Physics,
	Chinese Academy of Sciences, Beijing 100190, China}
 \affiliation{New Cornerstone Science Laboratory, 
	Beijing, 100190, China}

\author{Yi Lu}
\email{yilu@nju.edu.cn}
\affiliation{National Laboratory of Solid State Microstructures and Department of Physics, Nanjing University, Nanjing 210093, China}
\affiliation{Collaborative Innovation Center of Advanced Microstructures, Nanjing, 210093, China}

\author{Ke-Jin Zhou}
\email{kejin.zhou@diamond.ac.uk}
\affiliation{Diamond Light Source, Harwell Campus, Didcot OX11 0DE, UK}

\author{Donglai Feng}
\email{dlfeng@ustc.edu.cn}
\affiliation{National Synchrotron Radiation Laboratory and School of Nuclear Science and Technology, University of Science and Technology of China, Hefei, 230026, China}
\affiliation{New Cornerstone Science Laboratory, University of Science and Technology of China, Hefei, 230026, China}
\affiliation{Collaborative Innovation Center of Advanced Microstructures, Nanjing, 210093, China}

\date{\today}
\maketitle

\textbf{The striking discovery of high-temperature superconductivity (HTSC) of 80~K in a bilayer nickelate La$_3$Ni$_2$O$_7$ under a moderately high pressure of about 14 GPa ignited a new wave of studying HTSC in nickelates~\cite{WM327,PRL131-126001,Christiansson2023Correlated,interlayer_coup,2023arXiv230714819Z,2023arXiv230715276Z,2023arXiv230901148Y,Jun2023Emergence}. 
The properties of the parental phase at ambient pressure
may contain key information on basic interactions therein and bosons that may mediate pairing giving birth to superconductivity. 
Moreover, the bilayer structure of La$_3$Ni$_2$O$_7$ may suggest a distinct minimal model in comparison to cuprate superconductors. Here using X-ray absorption spectroscopy and resonant inelastic X-ray scattering, we studied  La$_3$Ni$_2$O$_7$ at ambient pressure, and found that  Ni 3$d_{x^2-y^2}$, Ni 3$d_{z^2}$, and ligand oxygen 2$p$ orbitals dominate the low-energy physics with a small charge-transfer energy. Remarkably, well-defined optical-like magnetic excitations were found to soften into a quasi-static spin-density-wave ordering, evidencing the strong electronic correlations and rich magnetic properties. Based on a Heisenberg spin model, we found that the inter-layer effective magnetic superexchange interaction is much larger than the intra-layer ones, and proposed two viable magnetic structures. Our results set the foundation for further exploration of La$_3$Ni$_2$O$_7$ superconductor. }

\vspace{8mm}

\noindent\textbf{Introduction}

Unlike cuprate superconductors, often characterized by a single Zhang-Rice singlet band consisting of Cu 3$d_{x^2-y^2}$ and O 2$p$ orbitals, multiple $d$ orbitals and Ni-O bilayer structure play critical roles in La$_3$Ni$_2$O$_7$~\cite{WM327, PRL131-126001, 2023arXiv231002915L, 2023arXiv231001952K,2023arXiv231105491C,interlayer_coup,Yang2023Inter}. In particular, the molecular bonding between the two Ni 3$d_{z^2}$ orbitals through the apical O $p_z$ orbital, together with Ni 3$d_{x^2-y^2}$ orbital, is widely established by theory and deemed as an essential ingredient for the low-energy electronic structure of La$_3$Ni$_2$O$_7$ \cite{2023arXiv230809698T,PRL131-126001,Yang2023Inter,interlayer_coup,2023arXiv231215727D,2023arXiv230901148Y,Lech2023Electronic}. 
However, the exact orbital occupancy and orbital character of La$_3$Ni$_2$O$_7$ remains elusive. If La$_3$Ni$_2$O$_7$ is viewed in close proximity to cuprates, $i.e.$, at the limit of a strong electronic correlation and a small charge-transfer energy, significant amount of electron holes would occupy the oxygen ligands, giving rise to Zhang-Rice-like physics~\cite{zhang1988effective}. On the other hand, supposing it was a sibling of infinite-layer nickelate superconductors, where the charge-transfer energy is rather comparable to the Coulomb repulsion, the participation of the oxygen ligands in the low-energy electronic structure would be much reduced~\cite{lin2021strong,shen2022role,hepting2020electronic}.

\begin{figure*}[htbp]
\centerline{\includegraphics[width=168mm,angle=0]{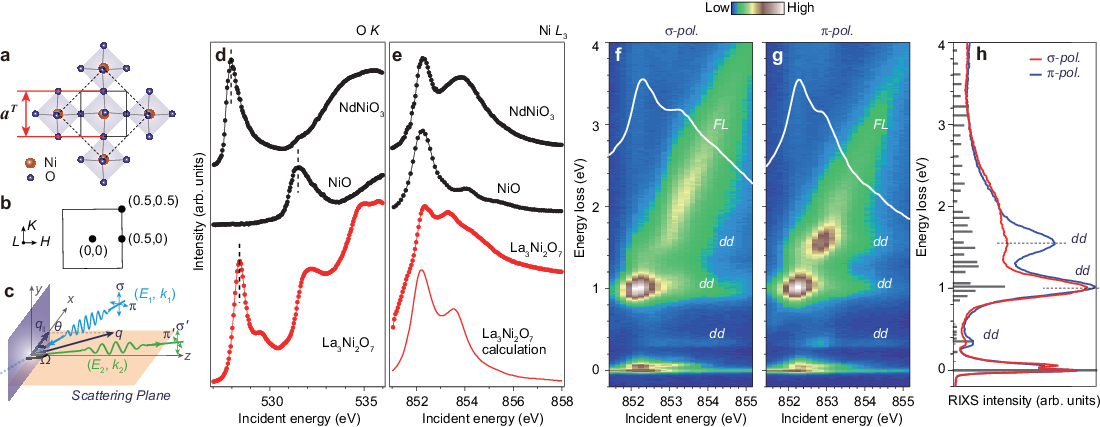}}
\caption{
\textbf{XAS and the incident energy dependent RIXS maps in La$_3$Ni$_2$O$_7$.}\label{fig:fig1}
\textbf{a,} Schematic top view of the NiO$_2$ plane in La$_{3}$Ni$_2$O$_7$. The solid black square represents the pseudo-tetragonal unit cell with a lattice constant $a^T$ $\sim$ 3.833 \AA, while the dashed black square represents the real orthorhombic in-plane unit cell when considering the tilting of Ni-O octahedra. 
\textbf{b,} In-plane Brillouin zone (BZ) for the pseudo-tetragonal unit cell.
\textbf{c,} Sketch of the RIXS experimental geometry. Details of the setup are described in Method. 
\textbf{d,e,} $\sigma$ polarised XAS spectra of La$_{3}$Ni$_2$O$_7$ (red filled circles) taken at the O $K$-edge (d) and Ni $L_3$-edge (e), respectively. The latter is displayed after subtracting the background of La $M_4$-edge. The calculated Ni $L_3$-XAS (red curve) is also displayed. XAS spectra measured on NiO (Ni$^{2+}$) and NdNiO$_3$ (Ni$^{3+}$) (black-filled circles) are shown as references.
\textbf{f,g,} RIXS intensity maps measured as a function of incident photon energy with $\sigma$- (f) and $\pi$-polarized photons. (g), respectively. The corresponding XAS spectrum is superimposed as a solid white curve on each map. Both XAS and RIXS spectra were collected at 20 K at a grazing-in incident angle of 20$^{\circ}$.
\textbf{h,} Integral of RIXS spectra in (f) and (g) over the incident energy range [851.8 eV, 853.4 eV]. The grey solid bars display the multiplet calculations for the Ni $L_3$ RIXS.
}
\label{Fig1}
\end{figure*}

The bilayer structure and the multi-orbital nature of La$_3$Ni$_2$O$_7$ have profound impact on its magnetism as well, which plays a pivotal role in theories on this novel superconductor, resembling the cuprate case~\cite{RevModPhys.75.913,WM327, interlayer_coup,inplane_coup}. Some suggest the importance of the interlayer antiferromagnetic coupling $J_z$ between $d_{z^2}$ orbitals~\cite{WM327, interlayer_coup}; 
some others advocate that the strong interlayer coupling would
cause the bilayer splitting of band structure, while in-plane magnetic exchange interactions play a dominant role in superconductivity~\cite{2023arXiv231105491C, 2023arXiv230809698T}. The intricate magnetic exchange may as well lead to other exotic phases with broken symmetries that have complex interplay with superconductivity, as seen in the cuprate superconductors~\cite{li2008unusual, keimer2015quantum}. In the as-grown La$_3$Ni$_2$O$_7$ crystal at ambient pressure, resistivity measurements have found a kink-like transition at around 153~K, which responds to the external out-of-plane magnetic field, implying a possible spin-density-wave (SDW) therein ~\cite{327_transport_SDW}. A recent $\mu$SR experiment on polycrystalline La$_3$Ni$_2$O$_7$ suggested that a static long-range magnetic order emerges below 148 K, consistent with an SDW internal field distribution~\cite{SL_uSR}. Moreover,
traces of a possible density wave have been discovered in a latest NMR report~\cite{NMR_327}. However, the exact nature of this density-wave state is still unclear. 

Given the currently limited knowledge on the essential electronic and magnetic energy scales, such as the charge-transfer gap and exchange interactions, experimental verification is indispensable. In this work, we employ X-ray absorption spectroscopy (XAS) and resonant inelastic X-ray scattering (RIXS) at both Ni $L_3$-edge and O $K$-edge of La$_3$Ni$_2$O$_7$ single crystal at ambient pressure. These spectroscopic and scattering techniques are sensitive to low-energy electronic and magnetic structures together with elementary excitations, and thus they are ideally suited for tackling the core issues in La$_3$Ni$_2$O$_7$.

\vspace{4mm}
\noindent\textbf{Electronic structure of La$_3$Ni$_2$O$_7$}

As-grown La$_3$Ni$_2$O$_7$ crystallizes in an orthorhombic structure with the space group of $Amam$~\cite{WM327}. We define the reciprocal space index $(H,K,L)$ based on the pseudo-tetragonal unit cell (Figs.~1a and 1b, Method).
Figure~1c shows the experimental geometry, in which the incident X-ray is linearly polarised, while the scattered X-ray is typically non-polarised but otherwise polarised if stated explicitly (see Method). 

Figures~1d and 1e illustrate XAS spectra of La$_3$Ni$_2$O$_7$ taken near the O $K$-edge and Ni $L_3$-edge, respectively. A sizable pre-edge peak at $\sim$ 528.5 eV is observed near the O $K$-edge, originating from oxygen 1$s$ electron excitations into the unoccupied oxygen 2$p$ ligand hole state near the Fermi level, as observed for the Zhang-Rice singlet state in cuprate superconductors~\cite{ChenCT1991PRL}. The Ni $L_3$-XAS data show a sharp resonant peak around 852.4 eV, followed by a broader satellite peak at a higher energy. As the Ni valence 2.5+ of La$_3$Ni$_2$O$_7$ falls in between the archetypal nickelates NiO and NdNiO$_3$, the XAS spectra of La$_3$Ni$_2$O$_7$ can be qualitatively understood in relation to these two. NiO resides in the charge-transfer regime of the Zaanen-Sawatzky-Allen scheme, whose large charge-transfer energy $\Delta$ ($\approx$ 5 eV) suppresses the charge fluctuations between the Ni 3$d$ and ligand oxygen 2$p$ orbitals despite their large orbital hopping integral~\cite{Sawatzky1984Magnitude}. Consequently, its ground state is well described by $\alpha|3d^8\rangle+\beta|3d^9\underline{L}\rangle$ ($\alpha^2+\beta^2 \lesssim 1$ and $\underline{L}$ denotes a ligand hole) with a dominant 3$d^8$ character ($\alpha^2\approx$ 0.8)~\cite{medarde1992r,mizokawa1995electronic,abbate2002electronic,horiba2007electronic}. On the other hand, the perovskite N d NiO$_3$ with a nominal 3$d^7$ configuration is widely acknowledged as a negative charge-transfer system, where electrons from ligand oxygen spontaneously transfer onto Ni cations, resulting in a ground state with a leading 3$d^8\underline{L}$ contribution~\cite{Green2016}. Such a substantial ligand hole concentration is underscored by the pronounced pre-edge hole peak in the O $K$-edge XAS of NdNiO$_3$, similar to that of La$_3$Ni$_2$O$_7$ (Fig.~1d). This is distinct from NiO, where the pre-peak is absent, and the unoccupied ligand states are at an elevated energy across the charge-transfer gap.
For the Ni $L_3$-XAS, the prominent resonant peak of La$_3$Ni$_2$O$_7$ is also observed for NiO and NdNiO$_3$ at a similar energy (Fig.~1e), which was previously identified as the Ni 2$p$ $\rightarrow$ 3$d^8$ or 3$d^8$ + 3$d^8\underline{L}$ $e_g$ transitions, respectively~\cite{vanderlaan1986, Green2016}. A broader satellite peak at a higher energy is likewise seen for NdNiO$_3$, originating mainly from a part of its ground state wavefunction that contains additional ligand holes~\cite{bisogni2016ground,Green2016,Lu2018}. 
The above comparison between the spectral features of La$_3$Ni$_2$O$_7$ and NiO/NdNiO$_3$ suggests a ground state with primarily 3$d^8$ occupancy on the Ni cation, accompanied by non-negligible ligand holes for the former.

Figures~1f and 1g show the incident-energy dependent RIXS measurements of La$_3$Ni$_2$O$_7$ across the Ni $L_3$-edge. A clear low-energy excitation ($\sim$ 70 meV) is observed near the elastic peak which will be discussed in the next section. The sharp XAS resonance at $\sim$ 852.4 eV decays mainly to a final state of a localized excitation at around 1 eV, known as the $t_{2g} \rightarrow e_g$ $dd$ orbital excitation similar to NiO and NdNiO$_3$~\cite{Chiuzbaian2005, Nag2020nio, bisogni2016ground}. The band-like fluorescence excitation, decaying from the broad satellite XAS peak, stems from the delocalized Ni-O hybridized continuum states~\cite{bisogni2016ground,Lu2018}. The intensity distribution of the fluorescence is more confined with $\pi$ polarization that couples stronger to the 3$d_{z^2}$ orbital, reflecting the more restricted electron kinetics perpendicular to the bilayer structure. It is noteworthy that, distinct from NdNiO$_3$, two extra $dd$ excitations show up in La$_3$Ni$_2$O$_7$ (at around 0.4 eV and 1.6 eV) and possess stronger intensities under $\pi$ polarization, suggestive of an enhancement along the 3$d_{z^2}$ orbital direction. 

To gain a quantitative understanding of XAS and RIXS measurements, we built a double-cluster model capturing the bilayer structure of La$_3$Ni$_2$O$_7$ and then carried out multiplet calculations for Ni $L_3$- XAS and RIXS spectra (see details in Section 2 of Supplementary Information). Systematic optimizations of the calculated spectra suggest that the charge-transfer energy $\Delta$ falls between 0 and 2 eV, pointing out the rather small-charge-transfer nature of La$_3$Ni$_2$O$_7$~\cite{chen2023critical}. This result is reasonable since $\Delta$ is $\sim$ 5 eV and $\sim$ 0 for NiO and NdNiO$_3$, respectively~\cite{Tanaka1994,Lu2018}. With the estimated range of $\Delta$, the ground state wavefunction of La$_3$Ni$_2$O$_7$ can be deduced to approximately $\alpha|3d^8\rangle+\beta|3d^8\underline{L}\rangle+\gamma|3d^7\rangle$ with leading $\alpha^2$ and $\beta^2$. The calculated XAS for $\Delta=$ 0.5 eV is shown in Fig.~1e, which corresponds to a ground state with $(\alpha^2, \beta^2, \gamma^2)$ $\approx$ (0.4, 0.3, 0.2). The corresponding RIXS calculation agrees well with the experiment, showing $dd$ excitations identified at comparable energies (Fig.~1h). Notably, we found that both the XAS line shape and the lower $dd$ excitation ($\sim$ 0.4 eV) in RIXS show marked difference upon tuning the inter-layer hopping strength mediated by the 3$d_{z^2}$ -$O_{AP}$ 2$p_z$ - 3$d_{z^2}$ orbital overlap in the calculation ($O_{AP}$ stands for the apical oxygen), underlining the importance of the inter-layer coupling for the electronic structure (Section 2 of Supplementary Information). This result is consistent with previous experimental report~\cite{WM327}, and lends support to several recent theoretical works emphasizing on the importance of the bilayer structure~\cite{kun_cpl,Yang2023Inter,qin2023high,arXiv:2307.16873,arXiv:2307.14965,interlayer_coup,
2023arXiv230809698T,PRL131-126001,Yang2023Inter,interlayer_coup,2023arXiv231215727D}.       

\begin{figure*}[htbp]
\centerline{\includegraphics[width=168mm,angle=0]{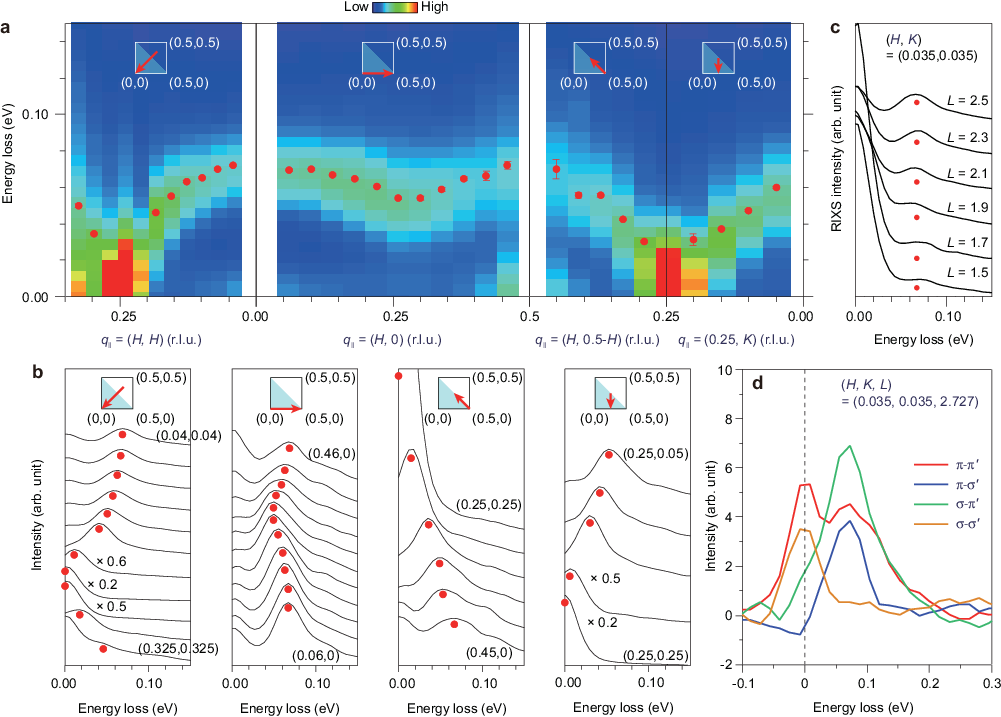}}
\caption{
\textbf{Energy-momentum dependent magnon in La$_3$Ni$_2$O$_7$.}
\textbf{a,} RIXS intensity maps along high-symmetry directions as indicated in the insets. Data were collected at 20 K using 852.4 eV, $\sigma$-polarised X-ray at the Ni $L_3$-edge. The red circles depict the peak positions of magnetic excitations here and throughout all panels of this figure. 
\textbf{b,} RIXS spectra at representative projected in-plane momentum transfers. The weaker excitations at $\sim$120 meV may result from the multi-magnons.
\textbf{c,} $L$ scan of RIXS spectra at at $q_\parallel$ = (0.035, 0.035).
\textbf{d,} Polarimetric RIXS data at $q$ = (0.035, 0.035, 2.727). The spectra are decomposed into $\pi-\pi^\prime$, $\pi-\sigma^\prime$, $\sigma-\sigma^\prime$ and $\sigma-\pi^\prime$ components.
}
\label{Fig4}
\end{figure*}

\vspace{4mm}
\noindent\textbf{Magnetic excitations}

Figure~2 summarises the detailed energy-momentum dependence of low-energy excitations in La$_3$Ni$_2$O$_7$ taken at the incident energy of 852.4 eV corresponding to the resonance of Ni $L_3$-XAS. Figures~2a and 2b show strongly dispersive excitations along directions illustrated in insets. The excitations reach maximal energy of about 70 meV at (0, 0) and (0.5, 0) while soften to zero energy (within the experimental energy resolution) at (0.25, 0.25), suggesting the presence of a quasi-static ordering. Similar excitations also appear when excited by $\pi$ incident X-rays polarisation (Fig.~S6). Along the out-of-plane direction, this mode does not exhibit sizable dispersion as a function of $L$, indicating its quasi-two-dimensional nature (Fig.~2c).

As both magnon and phonon excitations could appear in RIXS spectra, particularly within 100 meV that is closely relevant to both, the polarimetric RIXS was employed to analyze the outgoing X-rays linear polarisation for unraveling the origin of these excitations (see Methods). Clearly, as shown in Fig.~2d, the inelastic excitation is present under the $\pi-\pi'$, $\pi-\sigma'$, and $\sigma-\pi'$ channels, while gets much reduced under the $\sigma-\sigma'$ channel. 
Such behaviour is in excellent agreement with the assumption of a magnetic origin of the scattering~\cite{hill1996resonant}. Our multiplet RIXS calculation of magnetic excitations in the double-cluster model confirmed the outgoing polarisation dependence (Fig.~S5). Concerning phonons, in principle, their spectra weight should be present in the $\sigma-\sigma'$ channel. However, the corresponding polarimetric RIXS spectrum shows negligible spectral weight hence a minute contribution to the Ni $L_3$-RIXS (Fig.~2d).
We therefore conclude that the low-energy excitations observed at the Ni $L_3 $-edge are dominated by magnons. This immediately infers a quasi-static SDW ordering at (0.25, 0.25).  Remarkably, an SDW ordering was reported to exist near (0.25, 0.25) in the half-doped nickelate La$_{3/2}$Sr$_{1/2}$NiO$_4$ which has the same hole-doping level, $i.e.$, a nominal Ni$^{2.5+}$ valence state, as that of La$_3$Ni$_2$O$_7$ \cite{La3_2Sr1_2NiO4_INS}. In both cases, the magnon softens to almost zero energy near the SDW ordering wavevector, while their dispersions approaching $\Gamma$ point deviate drastically: there is an acoustic-like magnon in La$_{3/2}$Sr$_{1/2}$NiO$_4$, whereas it is absent here in Fig. 2a. 

By fitting the magnon spectra to a damped harmonic oscillator (DHO) function $\chi^{''}(q, \omega)$, we extracted the peak energy and width of the magnon (Section 4 of Supplementary Information)~\cite{peng2018dispersion}. 
Three possible spin configurations consistent with the spin order at $Q=(0.25, 0.25)$ can be constructed: 
the diagonal spin-charge stripe order as in half-doped La$_{3/2}$Sr$_{1/2}$NiO$_4$ where Ni$^{2+}$ spin and nominal Ni$^{3+}$ charge stripes intertwined (Stripe-1, Fig.~3a) ~\cite{La3_2Sr1_2NiO4_INS}; the SDW order could also be realised without the charge inhomogeneity, $i.e.$, a double-spin stripe order (Stripe-2, Fig.~3b); by exchanging the charge stripe positions, a third spin configuration could be achieved as a double spin-charge stripe order (Stripe-3 in Fig.~S10c. For all these SDW orders, owing to the strong bilayer bonding, spins are antiferromagnetically aligned in the top and bottom NiO$_2$ layers. To obtain the magnetic superexchange interaction parameters, we constructed an effective $J_1$-$J_2$-$J_z$ Heisenberg model:
$H=\sum_{i} J_z \vec{S}_i^{t}\cdot\vec{S}_i^{b}+\sum_{\langle ij\rangle\alpha} J_1 \vec{S}_i^{\alpha}\cdot\vec{S}_j^{\alpha}+\sum_{\langle\langle ij\rangle\rangle\alpha} J_2 \vec{S}_i^{\alpha}\cdot\vec{S}_j^{\alpha}$,
where $\alpha$ is the layer index for the bottom (b) or top (t) layer. $J_z$ is the inter-layer exchange coupling along the $c$-axis. $J_{1}$ and $J_{2}$ are the nearest-neighbor and next-nearest-neighbor exchange couplings, respectively, in a single Ni-O layer. The magnon dispersion within the linear spin wave theory was solved using the torque equation formalism~\cite{PhysRevB.70.064505} (Section 6 of Supplementary Information). We found that the magnon dispersion as well as its spectral weight distribution based on both Stripe-1 and Stripe-2 spin configurations agree with our RIXS result (Fig.~3). Owing to the scattering matrix effect, the simulated acoustic magnon spectra are significantly weaker than the optical magnon, consistent with the experimental findings.
~In general, the inter-layer effective superexchange interaction is an order of magnitude larger than that of the intra-layer. The finding of a dominant magnetic interaction along the molecular bonding direction is in good accordance with previous theoretical calculation~\cite{PRL131-126001}. Interestingly, the intra-layer effective superexchange interaction between next-nearest-neighbour spins on Ni$^{2+}$ sites shows comparable strength to that in the half-doped La$_{3/2}$Sr$_{1/2}$NiO$_4$ \cite{La3_2Sr1_2NiO4_INS}. 
Based on the above results and the currently limited information, 
we can conjecture the true spin configuration of La$_3$Ni$_2$O$_7$ is either Stripe-1 or Stripes-2, with the double spin stripe order just slightly lower in energy than the spin-charge stripe according to our calculations (see details in Section 5 of Supplementary Information).


\begin{figure*}[htbp]
\centerline{\includegraphics[width=160mm,angle=0]{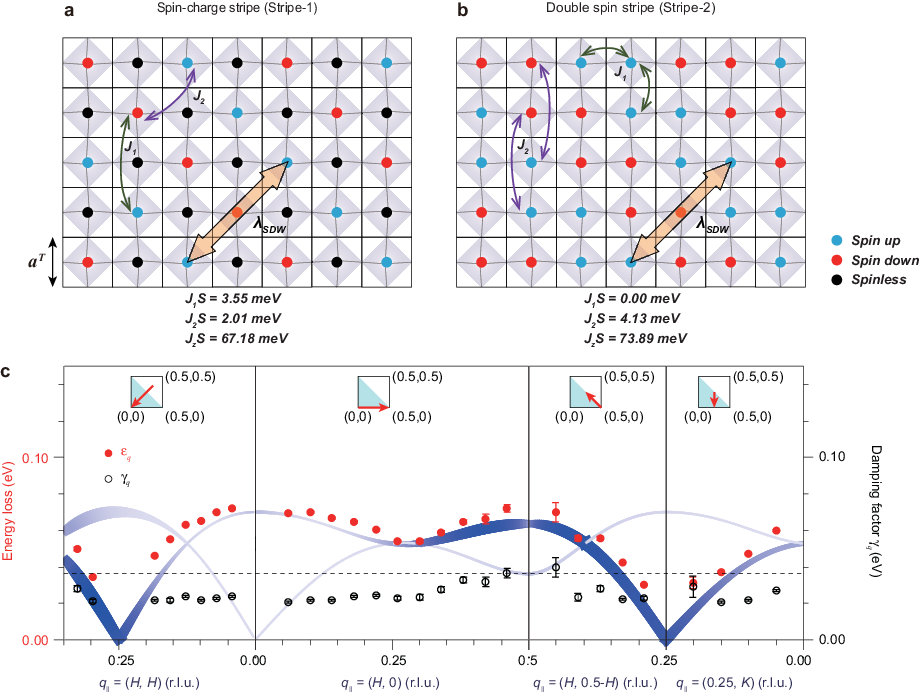}}
\caption{
\textbf{The spin configuration and the magnon dispersion of La$_3$Ni$_2$O$_7$}\label{fig:fig1}
\textbf{a,} The spin configurations for the spin-charge stripe order (Stripe-1). To simplify the sketch only nickel cations are shown. The blue, red and black circles represent spin up Ni$^{2+}$, spin down Ni$^{2+}$, and the nominal Ni$^{3+}$ sites, respectively. The solid lines illustrate the in-plane pseudo-tetragonal unit cells and the grey cubics represent the Ni-O octahedra. The fitted values of $J_1S$, $J_2S$, and $J_zS$ based on this spin configuration are noted (see details in Section 6 of Supplementary Information). \textbf{b,} The spin configuration for the double spin stripe (Stripe-2), and the fitted value of $J_1S$, $J_2S$, and $J_zS$. 
\textbf{c,} The experimental magnon dispersion $\epsilon_q$ (red filled circles) and damping factor $\gamma_q$ (black open circles) versus projected in-plane momentum transfer $q_\parallel$ along high-symmetry directions at 20 K. See fitting details in Section 4 of Supplementary Information. Error bars of $\epsilon_q$ were estimated by combining the uncertainty of the elastic peak position, linear background, and the standard deviation of the fits. Error bars of $\gamma_q$ were estimated by combining the standard deviation of the fits. The horizontal dashed line marks the total energy resolution (36 meV). Fittings for an effective $J_1$-$J_2$-$J_z$ Heisenberg model based on the Stripe-1 order are overlaid. The blue curves represent the dispersion of two magnon modes, where the thickness of the lines and the depth of their color represent the mode intensity. 
The detailed parameters are listed in Section 6 of Supplementary Information.
}
\label{Fig2}
\end{figure*}

\vspace{4mm}
\noindent\textbf{Spin-density-wave order}

We then took an explicit examination on the SDW order. Polarimetric RIXS was used to confirm the magnetic origin of low-energy excitations, likewise, it was applied to characterize this SDW order in La$_3$Ni$_2$O$_7$. Similar to the behaviour of magnons, the momentum-dependent quasi-elastic SDW scattering peak shows the same trend, $i.e.$, sizable scattering intensities under $\pi-\pi'$, $\pi-\sigma'$, and $\sigma-\pi'$ except for $\sigma-\sigma'$ (Figs. 4a and 4b), confirming the magnetic origin of such SDW order. Further insight into the nature of the SDW was gained through the energy dependence of the SDW scattering at its order wavevector across the Ni $L_3$-edge (Fig.~4c). Unlike the XAS spectra where La $M_4$ shows a greater absorption intensity than that of Ni $L_3$, the SDW scattering predominantly results from the Ni 3$d$ - O 2$p$ hybridised states. Furthermore, the SDW scattering peak exhibits a colossal polarisation dependence, namely, its intensity probed under $\pi$ polarisation is about 30 times higher than that with $\sigma$ polarisation. Figure~4d gives an example taken with 852.4~eV photons, which may indicate its strong association with Ni 3$d_{z^2}$ orbital. The half-width at half-maximum $\Gamma$ = 0.0022$\pm$0.0002 r.l.u. of the scattering peak corresponds to a relatively short in-plane correlation length ($\xi_H$ = 1/$\Gamma$) of $\sim$27.7 nm. A much broader peak is observed as a function of $L$ along the direction of (0.25, 0.25, $L$) establishing the quasi-two-dimensional nature of such SDW order (Fig.~4e). 

The temperature dependence of the SDW order illustrates a substantial reduction in both the intensity and the correlation length when the temperature is raised above $\sim$150~K (Figs.~4f-4h). 
While the SDW wavevector does not exhibit a discernible temperature dependence (Fig.~4i).
The discovery of the SDW with a characteristic temperature of around 150~K agrees well with previous transport, NMR and $\mu$SR measurements on La$_3$Ni$_2$O$_7$~\cite{327_transport_SDW,NMR_327, SL_uSR}.

\begin{figure*}[t!]
\centerline{\includegraphics[width=168 mm,angle=0]{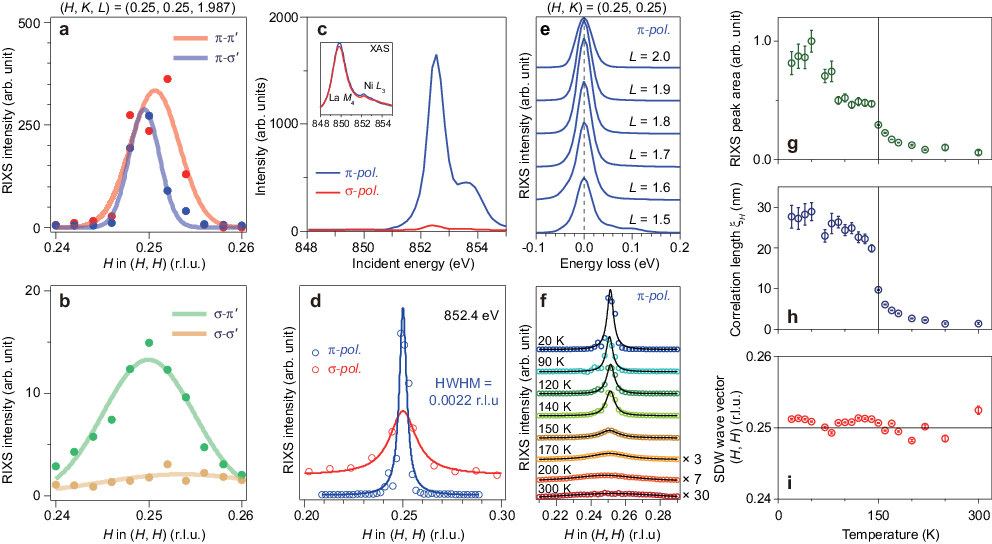}}
\caption{
\textbf{SDW order at (0.25, 0.25) of La$_3$Ni$_2$O$_7$. }
\textbf{a,b} Polarimetric RIXS data. The spectra are decomposed into $\pi\pi^\prime$, $\pi\sigma^\prime$, $\sigma\sigma^\prime$ and $\sigma\pi^\prime$ components.
\textbf{c,} SDW peaks intensities as a function of incident photon energy and polarization. The inset shows the XAS spectra at the La $M_4$-edge and the Ni $L_3$-edge.
\textbf{d,} SDW peak intensity integrated over an energy window of 36.5 meV (the total energy resolution) as a function of projected momentum transfer ($q_\parallel$) along the ($H$, $H$) direction. 
The fitted peak center values are $\sim$ 0.25 r.l.u. and the corresponding half width at the half maximum (HWHM) is 0.0022 r.l.u.
\textbf{e,} $L$ scan of RIXS spectra at $q_\parallel$ = (0.25,0.25).
\textbf{f,} SDW peaks and their Lorentzian fits along the ($H$, $H$) direction at various temperatures.
\textbf{g-i,} Temperature dependence of the SDW peak area (g), the correlation length (h) and the SDW wave vector position (i).
}
\label{Fig6}
\end{figure*}

\vspace{4mm}
\noindent\textbf{Discussion and Conclusion}

Our RIXS and XAS measurements revealed the dispersive magnon and SDW order below 150 K in La$_3$Ni$_2$O$_7$. Detailed analysis suggests that Ni 3$d_{x^2-y^2}$, Ni 3$d_{z^2}$, and O 2$p$ orbitals dominate the low-energy physics with charge-transfer energy less than 2~eV, and the inter-layer effective magnetic superexchange interaction is much larger than the intra-layer ones. These give critical information for constructing the minimal model for La$_3$Ni$_2$O$_7$ superconductor. 

Apart from the extraordinary bilayer structure and the associated predominant magnetic exchange interaction, the electronic structure of La$_3$Ni$_2$O$_7$ fits in general into the family of Ruddlesden-Popper (RP) nickelates. The formation of the Zhang-Rice-like hole band, the small charge-transfer energy, and the well-defined dispersive magnon allude to its nature of the strong electronic correlations ~\cite{liu2023electronic}. The above are typical characteristics of the strongly correlated cuprates where charge- and spin-density modulation can take place. Moreover, the occurrence of SDW order at (0.25, 0.25) is reminiscent of that in the half-doped single-layer La$_{3/2}$Sr$_{1/2}$NiO$_4$, where a spin-charge stripe order exists, and implies the tendency to a charge-density-wave instability in La$_3$Ni$_2$O$_7$ as illustrated in the scenario of Stripe-1 (Fig.3a)~\cite{La3_2Sr1_2NiO4_INS,chen2023critical}. Indeed in layered half-doped RP nickelates, manganites, and cobaltates, the spin-charge intertwined order is prevailing~\cite{La3_2Sr1_2NiO4_INS,sternlieb1996charge,zaliznyak2000independent}. On the other hand, the double spin stripe order accommodating homogeneous charge-density (Stripe-2, Fig.~3b) may be possible too as the 3$d_{x^2-y^2}$ orbitals are more itinerant in-plane than the 3$d_{z^2}$ orbitals. This is similar to the collinear double spin stripe state in the correlated FeTe~\cite{fete_spinwave}.
Verifying the existence of the charge-density-wave order could unequivocally clarify the spin configuration of La$_3$Ni$_2$O$_7$.

Finally, we would like to extrapolate our findings to superconducting La$_3$Ni$_2$O$_7$ under a moderately high pressure: here, a structural phase transition accompanied by a few percent shrinkage of the lattice constants, and the Ni-O-Ni bonding angles between adjacent NiO$_6$ octahedra straighten to 180$^\circ$~\cite{WM327}.
Consequently, the electronic hopping is likely to increase, potentially suppressing density waves that compete with the superconductivity~\cite{Souliou2018Pressure, Taillefer2018Pressure}. Furthermore, the magnetic superexchange $J_z$ may get significantly enlarged due to the increased hopping along Ni-O$_{AP}$-Ni, highlighting the vital role of the inter-layer AFM interaction in the HTSC of such a bilayer nickelate.  
~\\

\begin{acknowledgments}

This work was supported by National Natural Science Foundation of China (Nos.11888101, U2032208, 12274207, 12174428, and 12174454), the New Cornerstone Science Foundation, and the National Key R\&D Program of China (Nos. 2023YFA1406304, 2022YFA1403000, 2023YFA1406500, and 2023YFA1406002). We acknowledge Diamond Light Source for providing beamtime at I21 Beamline under Proposal MM35805 and the science commissioning beamtime for using the polarimeter at I21. Work at SYSU was as well supported by the Guangdong Basic and Applied Basic Research Funds (No. 2021B1515120015), Guangzhou Basic and Applied Basic Research Funds (Nos. 202201011123, 2024A04J6417), and Guangdong Provincial Key Laboratory of Magnetoelectric Physics and Devices (No. 2022B1212010008). 

\end{acknowledgments}


~\\
\noindent\textbf{\large Methods}~\\
\noindent\textbf{Sample fabrication}: 
La$_3$Ni$_2$O$_7$ sample was fabricated by the high oxygen pressure floating zone technique and the details are described in \cite{327_transport_SDW}. The sample quality was checked by X-ray diffraction (XRD) and Laue diffraction (see details in Fig. S1). 
Samples were cleaved to get a flat, clean surface before RIXS measurements.~\\

\noindent\textbf{XAS and RIXS measurements}: 
XAS and RIXS measurements were performed at Beamline I21 at Diamond Light Source~\cite{zhou2022i21}. In this work, we describe the structural properties of
La$_3$Ni$_2$O$_7$ referencing to a pseudo-tetragonal unit cell with cell
parameters $a^T$ = $b^T$ $\sim$ 3.833 Å and $c$ = 20.45 Å. Reciprocal lattice units (r.l.u.) are defined (where $2\pi/a^T=2\pi/b^T=2\pi/c=1$) with $\mathbf{Q}=H\mathbf{a}^{T*}+K\textbf{b}^{T*}+L\textbf{c}^{*}$. The crystallographic $a^T$–$c$ ($b^T$–$c$) plane of La$_3$Ni$_2$O$_7$ single crystal was aligned within the horizontal scattering plane (Fig.~1c). The polar angular offsets ($\theta$ and $\chi$) of the crystal were aligned by the (002) diffraction peak, and the azimuthal offset ($\phi$) by SDW order peak, such that the $c^*$ axis lays in the scattering plane. The spectrometer arm was at a fixed position of $\Omega=154^\circ$ except for $L$ scans where variable $\Omega$ was employed.

XAS spectra were collected with a grazing incidence angle of $\theta_0 = 20^\circ$ to probe both in-plane and out-of-plane unoccupied states. All XAS measurements were done at a temperature of 20\;K with the exit slit opening to 30 $\mu$m. Total electron yield XAS spectra were collected using the draincurrent and normalised to the incoming beam intensity. Both linear vertical ($\sigma$) and horizontal ($\pi$) polarisations were used.

Energy-dependent RIXS measurements were performed at the grazing incidence angle of $\theta_0 = 20^\circ$ and the temperature of 20\;K. The exit slit was open to 30 $\mu$m corresponding to an average energy resolution of 40 meV (FWHM). The incident energy range went from 851 to 855\;eV in steps of 0.2\;eV to fully capture the resonance behaviour across the Ni-$L_3$ absorption peaks. 

Momentum-dependent RIXS measurements were performed at the resonant energy of 852.4 eV at a temperature of 20\;K with the exit slit opening to 20 $\mu$m corresponding to an average energy resolution of 36 meV (FWHM). RIXS spectra were collected using both $\sigma$ and $\pi$ polarisations. The grazing out geometry ($\theta > \Omega/2$) was applied for the acquisition of RIXS spectra shown in the main text. 

Polarimetric RIXS apparatus employs a graded multilayer designed for the Ni $L_3$-edge with a grazing incidence angle of 20$^\circ$ lying perpendicular to the scattering plane. Measurements were performed at Q = (0.035, 0.035, $L$) and around (0.25, 0.25, $L$) to analyse the outgoing X-rays linear polarisation of the magnon and SDW ordering, respectively. The total energy resolution of the polarimetric RIXS is $\sim$ 55 meV (FWHM). Since the multilayer does not work at the exact Brewster's angle, the outgoing polarised RIXS (the indirect RIXS) from the reflection of the multilayer will be a mixture of linearly polarised spectra. The direct and indirect RIXS spectral intensities are then given by the following formula:
\begin{equation}
I_{direct} = I_{\sigma'} + I_{\pi'}
\end{equation}
\begin{equation}
I_{indirect} = R_{\sigma'}I_{\sigma'} + R_{\pi'}I_{\pi'}     
\end{equation}
where $I_{direct}$ and $I_{indirect}$ stands for the outgoing nonpolarised and mixed polarised RIXS spectral intensity, respectively. From the above formula, the outgoing $\sigma'$ and $\pi'$ polarised RIXS spectra can be deduced: 
\begin{equation}
I_{\pi'} = \frac{I_{indirect} - R_{\sigma'}I_{direct}}{R_{\pi'} - R_{\sigma'}}
\end{equation}
\begin{equation}
I_{\sigma'} = \frac{I_{indirect} - R_{\pi'}I_{direct}}{R_{\sigma'} - R_{\pi'}}
\end{equation}
In the above, R$_{\sigma'}$ (R$_{\pi'}$) refers to the multilayer reflectivity of the outgoing $\sigma'$ ($\pi'$) polarised X-ray photon. At the Ni $L_3$-edge, R$_{\sigma'}$ and R$_{\pi'}$ is 14.1\% and 9.1\%, respectively, based on the calibration of the multilayer.

~\\
\noindent\textbf{Theoretical calculations}:
The Ni $L_3$-edge XAS and RIXS calculations shown in Figure 1 were performed employing a fully correlated Ni$_2$O$_{11}$ cluster model, accounting for the two corner-sharing NiO$_6$ octahedra within the pseudo-tetragonal unit cell. The noninteracting part of the Hamiltonian integrates material-specific on-site energies and hybridizations involving Ni 3$d$ and O 2$p$ orbitals, along with spin-orbit coupling within the Ni core 2$p$ and 3$d$ shells. Full Coulomb interactions within the Ni 3$d$ shell and between the Ni 2$p$ and 3$d$ shells are included, with parametrization by Slater integrals scaled at 0.8 based on atomic Hartree-Fock values~\cite{cowan1981}. Comprehensive details regarding model construction and relevant parameters are described in Section 2 of Supplementary Information. The model was solved using the exact diagonalization method as implemented in \textsc{Quanty}~\cite{quanty}.

The DFT calculations employ the Vienna ab-initio simulation package (VASP) code \cite{kresse1996efficient} with the projector augmented wave (PAW) method \cite{kresse1999ultrasoft}. The Perdew-Burke-Ernzerhof (PBE)  exchange-correlation functional \cite{perdew1996generalized} is used. The energy cutoff energy for expanding the wave functions into a plane-wave basis is set to be 500 eV. The $\Gamma$-centered k-mesh is used in KPOINTS files which are generated by VASPKIT \cite{wang2021vaspkit} with the KPT-resolved value equal to 0.02 for different unit cells. 
The SDW orders are calculated using the simplified rotation invariant approach based on the DFT+U method introduced by Dudarev $et$ $al.$ \cite{dudarev1998electron}. Then, the effective Heisenberg interactions for the SDW orders are constructed. The magnon dispersion within the linear spin wave theory are calculated using the torque equation formalism\cite{PhysRevB.70.064505,lin2021strong}. The RIXS intensity for the magnon mode in the $\sigma$-$\pi$ polarization channel is calculated following the reference \cite{PhysRevLett.105.167404}. 
More details can be found in Sections 5 and 6 of Supplementary Information.
~\\

\noindent\textbf{\large Competing interests}~\\Authors declare that they have no competing interests.

~\\
\noindent\textbf{\large Data availability}~\\
All data are available in the main text and Supplementary Information.


\end{document}